# NMR crystallography reveals carbonate induced Al-ordering in ZnAl layered double hydroxide

Sambhu Radhakrishnan,[a,b,†] Karl Lauwers,[b,†] C. Vinod Chandran,[a,b] Julien Trebosc,[c] Johan A. Martens,[b] Francis Taulelle,[b] Christine E. A. Kirschhock[b] and Eric Breynaert[a,b,*]

[a] Dr. Sambhu Radhakrishnan, Dr. Vinodchandran C., Dr. Eric Breynaert
NMRCoRe,
KU Leuven
Celestijnenlaan 200F Box 2461, B-3001, Belgium.
E-mail: Eric.Breyanaert@kuleuven.be

[b] Dr. Sambhu Radhakrishnan, Dr. Karl Lauwers, Dr. Vinodchandran C., Prof. Johan A. Martens. Prof. Francis Taulelle, Prof. Christine E. A. Kirschhock, Dr. Eric Breynaert
Centre for Surface Chemistry and Catalysis - Characterization and Application Team (COK-KAT)
KU Leuven
Celestijnenlaan 200F Box 2461, B-3001, Belgium.

[c] Dr. Julien Trebosc
Institut Chevreul CNRS FR-2638
Université de Lille, Bât C7, ENSCL, ave Mendeleiev 59652 Villeneuve d'Ascq, France.

Supporting information for this article is given via a link at the end of the document.

**Abstract:** Layered double hydroxides (LDHs) serve a score of applications in catalysis, drug delivery, and environmental remediation. Smarter crystallography, combining X-ray diffraction and NMR spectroscopy revealed how interplay between carbonate and pH determines the LDH structure and Al ordering in ZnAl LDH. Carbonate intercalated ZnAl LDHs were synthesized at different pH (pH 8.5, pH 10.0, pH 12.5) with a Zn/Al ratio of 2, without subsequent hydrothermal treatment to avoid extensive recrystallisation. In ideal configuration, all Al cations should be part of the LDH and be coordinated with 6 Zn atoms, but NMR revealed two different Al local environments were present in all samples in a ratio dependent on synthesis pH. NMR-crystallography, integrating NMR spectroscopy and X-ray diffraction, succeeded to identify them as Al residing in the highly ordered crystalline phase, next to Al in disordered material. With increasing synthesis pH, crystallinity increased and the sidephase fraction decreased. Using $^1$H-$^{13}$C, $^{13}$C-$^{27}$Al HETCOR NMR in combination with $^{27}$Al MQMAS, $^{27}$Al-DQ-SQ measurements and Rietveld refinement on high-resolution PXRD data, the extreme anion exchange selectivity of these LDHs for $CO_3^{2-}$ over $HCO_3^-$ was linked to strict Al and $CO_3^{2-}$ ordering in the crystalline LDH. Even upon equilibration of the LDH in pure $NaHCO_3$ solutions, only $CO_3^{2-}$ was adsorbed by the LDH. This reveals the structure directing role of bivalent cations such as $CO_3^{2-}$ during crystallisation of $[M^{2+}_4M^{3+}_2(OH)_2]^{2+}[A^{2-}]_1 \cdot yH_2O$ LDH phases.

## Introduction

Layered double hydroxides (LDHs), also called anionic clays, are lamellar materials, isostructural to the natural mineral hydrotalcite. Isomorphic substitution of di- by tri-valent metal cations in the brucite structure results in a net positive charge of the octahedral layers. This positive charge is compensated by anions located in hydrated interlayers. LDHs are represented by the general formula $[M^{2+}_{1-x}M^{3+}_x(OH)_2]^{x+}[A^{n-}]_{x/n} \cdot yH_2O$, where $M^{2+}$ ($Mg^{2+}$, $Ni^{2+}$, $Zn^{2+}$, etc.), and $M^{3+}$ ($Al^{3+}$, $Fe^{3+}$, etc.). x denotes the fraction of isomorphic substitution $M^{3+}/(M^{2+} + M^{3+})$, while $A^{n-}$ ($CO_3^{2-}$, $NO_3^-$, $Cl^-$, etc.) is the charge compensating interlayer anion.[1] As result of the high variability in cation and anion composition, LDHs are promising materials for use as anion exchanger, adsorbent, catalyst, and for controlled release of pharmaceutical components.[2] Recently, even mesoporous luminescent, and mixed ligand luminescent LDHs have been reported.[3,4] LDH structures are typically described in a rhombohedral or hexagonal crystal symmetry with three or two layers in a unit cell in the hexagonal crystal system. Usually, cation-disorder is assumed, treating cation sites of tri- and divalent ions as equivalent.[5,6] However, Kamath and coworkers emphasized anomalies and discrepancies of rhombohedral LDH structure refinement described by disordered cation models.[7,8] Significantly different ionic radii of di- and trivalent cations prevent the description of corresponding LDH structures with a single metal-oxygen bond length in rhombohedral topology with ABC layer stacking. An LDH model for highly periodic structures should therefore include distinct crystallographic sites for the di- and trivalent cations, instead of identical sites for all cations. In addition, the metal ion distribution should obey Pauling's rule, avoiding small, trivalent cations, specifically Al cations, to occupy edge-sharing octahedra. Reappraisal of the LDH polytype symmetry was established when LDH structures were refined by the Rietveld method in a structure model with ordered cations in a monoclinic representation of the averaging rhombohedral setting.[9] The suggested one-layer cell in the space group C2/m focusing on the organization of heavy Zn and lighter Al atoms in the layers but did not take into account any possible ordering of the interlayer species like anions and water, owing to the lack of corresponding reflections.

Local order has recently been investigated in more detail with nuclear magnetic resonance (NMR). Solid state $^1$H and $^{27}$Al NMR spectroscopy offer complementary information on local defects and Al-rich phases in LDH structures. A fully ordered ZnAl LDH with Zn/Al ratio of 2 for instance, should exhibit a single $^{27}$Al NMR resonance corresponding to every $Al^{3+}$-ion surrounded by six Zn cations.[10] However, even highly crystalline LDH materials exhibited additional Al resonances, indicative for the presence of local defects, structural disorder, or presence of side phases.[10] The relative concentration of the different Al species, as well as the fraction of perfectly ordered Al centers can be calculated from $^{27}$Al NMR.[11] Also fast-spinning $^1$H MAS NMR spectroscopy has been used to elucidate cation ordering in LDH materials.[12] Neglecting interlayer interactions and based on a purely statistical cation distribution, ZnAl LDH structures potentially can only contain three local hydroxyl environments within the layers (e.g., $Zn_3$-OH, $Zn_2Al$-OH, and $ZnAl_2$-OH hydroxyl groups). $^1$H NMR spectra can therefore provide an indication of the metal ion distribution, since perfectly ordered ZnAl LDHs with Zn/Al ratio 2 should exclusively exhibit signatures of hydroxyl functions connected to exactly one Al, and two Zn atoms.[10] Though the complete correlation of $^1$H NMR with cation disorder is still debated, hitherto all contradictory results have been ascribed to artifacts resulting from synthesis methods inducing various degrees of disorder and defects.[13,14] Carbon species in the interlayers of carbonate-intercalated ZnAl LDHs (ZnAl-$CO_3$) have been investigated with NMR spectroscopy, revealing two types of $^{13}$C signals assigned to either exchangeable carbonate and hydrogen carbonate species or to two carbon species interacting with the LDH layer or with impurities present.[15–17] Unambiguous assignment of the $^{13}$C NMR signals observed in carbonate intercalated LDHs also remains a topic of debate.

LDH materials are typically synthesized by co-precipitation in a pH stated medium at pH between 8 and 10.[18] Upon mixing, the chemicals, nearly instantly form amorphous hydroxides. With time, these amorphous hydroxides transform into crystalline layered double hydroxides, a process dependent on temperature, pH and the identity of anions and cations in the system.[19] As previously shown by Kloprogge et al., synthesis pH is a deterministic parameter for the final structure of the LDH: crystallinity and temperature stability of ZnAl LDH improve with increasing synthesis pH.[20] In addition to evaluating the impact of pH on the crystallinity of the product formed at low (60°C) temperature, this work aims to highlight the structure directing role and more specifically the charge ordering function of carbonate anions on ZnAl LDH, disentangling this effect from the impact of temperature and pH. ZnAl-$CO_3$ LDH materials were synthesized by co-precipitation at pH 8.5, pH 10.0 and pH 12.5. Crystal structure, crystallinity and morphology of the LDH phases were characterized with PXRD and SEM, respectively, while local structural differences were studied in detail with solid state NMR spectroscopy. The symmetry and full structural models, including Al as well as anion ordering, were determined using NMR crystallography. Combining X-ray diffraction with Rietveld refinement and advanced solid state NMR, diffraction enables to selectively derive the structure and periodicity of the crystalline parts of a sample. NMR reveals local structural information on all phases present and exploits advanced experiments and filtering schemes, to highlight specific parts in the sample or selectively investigate well defined interactions between atoms.

## Results and Discussion

LDH synthesis and basic characterization.

Crystalline LDH typically exhibit a hexagonal platelet-like morphology and can easily be discerned by Scanning Electron Microscopy (SEM), as shown in Figure 1 for LDH125. LDH125 consists almost exclusively of uniform hexagonal platelets of about 500 nm diameter, aggregated in chunks of 10 to 20 µm. In LDH85, an extensively disordered (most likely amorphous) phase is observed in addition to the platelets, which are not as well defined, showing cavities and crumbled edges. Efforts to perform elemental analysis (EDS) on the disorder phase were unreliable due to a combination of beam induced damage at high accelerating voltages (~20 kV) and to the inaccuracy associated with the large interaction volume of the beam with the close proximity of the phases. The morphology of LDH100 is similar to that of LDH125, although the platelet size varies more strongly. The XRD patterns of the ZnAl-$CO_3$ LDHs indicate a similar trend as the SEM characterization (Figure 2). Comparison of the diffractograms reveals LDH125 as the most crystalline phase, exhibiting a baseline with least intensity and the sharpest reflections, even clearly showing intra-layer signals. In the series LDH125, LDH100 and LDH85, especially LDH85 exhibited broad reflections and unresolved intensity around 20 and 27 degrees 2θ. The increase in crystallinity with increasing synthesis pH is consistent with the trend observed by Kloprogge et al. between pH 6.0 and 11.5.[20] Table 1 summarizes different material characteristics and lattice parameters of the series. Increasing the synthesis pH from 8.5 to 10 and 12.5, the first reflection in the diffractograms shifts to the left in LDH125 and LDH100 as compared to LDH85, corresponding to higher interlayer distances and hence a larger $c$ lattice constant. Also the in-plane lattice constants ($a$ and $b$) increased as compared to LDH85. In ZnAl LDHs, the in-plane lattice constants typically increase with higher Zn/Al ratio, due to increased presence of Zn atoms with larger ionic radius compared to Al.[1,5] ICP results indicate bulk Zn/Al ratios of 2 for LDH125 and LDH100, but show a slightly too high Al content for LDH85. Dosing $Zn(NO_3)_2$ and $Al(NO_3)_2$ to alkaline aqueous solution containing carbonates, result in an initial, very fast (< 10 min), co-precipitation of amorphous mixed hydroxide nanoparticles reflecting the metal cation composition of the solution the metals were dosed from.[19] Subsequently these mixed hydroxide amorphous phases convert to layered double hydroxides. The slightly higher Al content of the LDH85 solid as compared to LDH100 and LDH125 is attributed to co-precipitation of amorphous $Al(OH)_3$ phases exhibiting a much lesser solubility at lower pH as compared to $Zn(OH)_2$.[19,21,22]

The high order and strict ratio of Zn/Al of 2 manifests in unit cells with lower symmetry than R-3m (vide infra). Nonetheless, in the found monoclinic unit cells the average metal-metal distances, which correspond to 1/3 of the $b$-lattice constants, are in full accordance of a Zn/Al ratio of 2/1 for LDH125 and LDH100 (Table 1).[23] As expected for these minerals, the BET surface area determined at liquid nitrogen temperatures is relatively low, being the highest in LDH85 as consequence of smaller particle size and significant presence of disordered phases.

The results from XRD, SEM and $N_2$ physisorption are corroborated by $^{27}Al$ MAS NMR. 1D $^{27}Al$ NMR reveals the samples exclusively contain 6-coordinated Al, present in at least 2 Al environments with a ratio dependent on the synthesis pH (Figure 3; Figure S-1). It must be noted that these ratio's cannot be used to derive the Zn/Al ratio of the LDH phases, since both Zn and Al were initially co-precipitated as amorphous mixed hydroxides with a Zn/Al ≈ 2.[19] Combining $^{27}Al$ MQMAS (Figure S-2) with $^{27}Al$ direct excitation spectra (1D) has been demonstrated to enable reliable phase identification in transition extraction of major components and their ratio's (Table 2, Figure S1). The sharp component with isotropic $^{27}Al$ chemical shift at ca. 14.6 ppm and $C_Q$ of 0.73 MHz, corresponds to a symmetric local environment of Al surrounded by 6 Zn atoms (Al-1),[10] and gains intensity with increasing synthesis pH. The broad resonance with $^{27}Al$ isotropic chemical shifts between 11 and 14 ppm and $C_Q$ values exceeding 1.5 MHz corresponds to Al species occurring in a distorted 6-coordinated environment (Al-2). Decomposition of the $^{27}Al$ spectra required using the Czjzek model (DMFIT) to account for both the chemical shift and quadrupole parameter distribution.[24,25] The fits indicated higher chemical shift distributions were present for Al-2 as compared to Al-1, owing to a higher degree of disorder in Al-2. The ratio Al-1/Al-2 increases with increasing synthesis pH, indicating a correlation to crystal morphology and crystallinity and/or to the presence of amorphous (non – LDH) phases. Based on this, it can be derived that Al-1 corresponds to the ordered LDH fraction. $^1H$ decoupled $^{13}C\{^1H\}$ CPMAS spectra (Figure 4) indicate the carbonate speciation of all samples is nearly predominantly $CO_3^{2-}$, exhibiting chemical shifts in the range of 169 to 172 ppm. The spectra for LDH125 and LDH100 show a single, though asymmetric peak envelope with maxima at 171 and 171.8 ppm, respectively. In contrast, LDH85 exhibits two resolved envelopes with chemical shifts around 171.8 and 169.5 ppm. This supports the difference between LDH85 and the other two samples, observed with chemical analysis. LDH85 exhibits a C/Al ratio slightly higher than 0.5, while the ratio for LDH100 and LDH125 is strictly 1:2, the ideal value required for a fully $CO_3^{2-}$ exchanged and ordered Zn/Al LDH. $^1H$-$^{13}C$ CPMAS spectra acquired with different contact times (Figure S-5, Supporting Info) indicate a different buildup behavior for the two envelopes observed in the $^{13}C$ spectra for LDH85, implying they occur in different, non-interacting and separate regions of the sample. The $^{13}$C signal at 171.8 ppm exhibited a buildup up to a CP contact time of 1.25 ms followed by a decay of the signal. This behavior is similar to the corresponding signal at 171 ppm in sample LDH125 (Figure S-6), indicating this carbonate species to be present in the crystalline LDH fraction. The resonance at 169.5 ppm in LDH85 keeps building up magnetization up to 5.5 ms and decays at longer contact times, confirming presence of different alumina phases, also in the environment providing dipolar coupling to a different $^1$H spin pool.[26] The combination of these results, implies the increased C/Al ratio observed for LDH85 correlates to Al either in non-LDH side-phases, or in severely distorted, less ordered LDH regions. Furthermore, these results clearly indicate the samples all contain a highly ordered LDH phase next to fractions of disordered side-phases or LDH material without long- or short-range order. A more detailed view should be attainable by NMR crystallography, elucidating the overall structure of the sample by combining X-ray diffraction to reveal the structure of the ordered fraction and advanced multidimensional NMR spectroscopy to obtain local structural information also for the disorder phases.

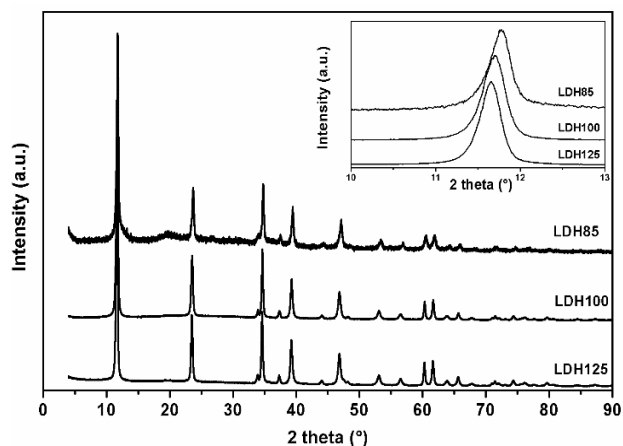

**Figure 2**: XRD patterns of ZnAl-CO3 LDHs. Decreasing linewidths illustrate an increase in crystallinity with increasing synthesis pH. The inset shows the first reflection position in detail.

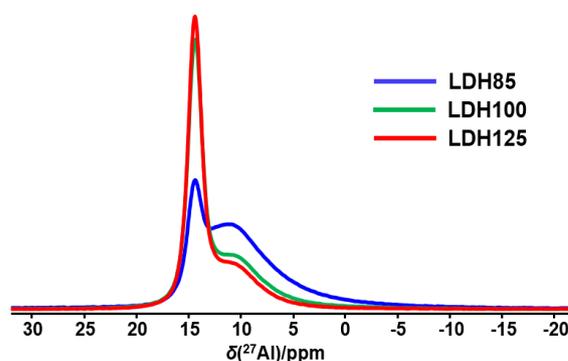

**Figure 3**: $^1H$ decoupled $^{27}Al$ NMR spectra of ZnAl-CO3 LDHs synthesized at different pH recorded at 11.4 T.

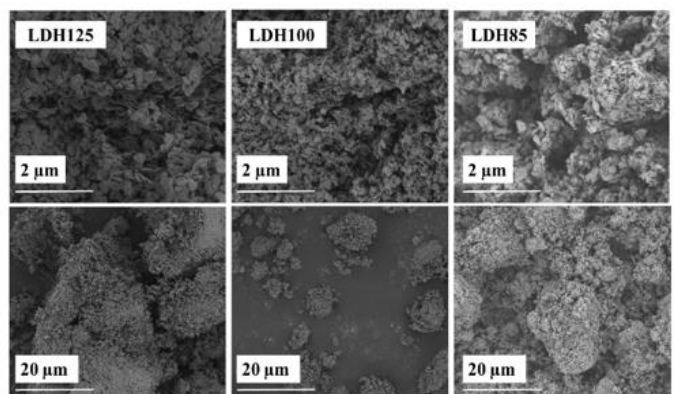

**Figure 1**: SEM images of ZnAl-CO3 LDHs. Clear hexagonal platelets are formed in LDH125, whereas more disorder is present in LDH85. LDH100 forms uniform platelets although the size is more varied.

Table 1. Table 1. Zn/Al ratio, carbon and water content, BET surface areas, refinement results and lattice parameters of ZnAl-CO3 LDHs synthesized at different pH.

| ZnAl-CO$_3$ | Zn/Al | C/Al | Ow/Al | BET surface area (m$^2$/g) | Space group | R-values (%) | | | a (Å) | b (Å) | c (Å) | β (°) |
|---|---|---|---|---|---|---|---|---|---|---|---|---|
| | | | | | | Rp | Rwp | RF$^2$ | | | | |
| LDH125 | 2.0 | 0.5 | 1.5 | 23.8 | P2/n | 2.9 | 3.9 | 9.8 | 5.317 | 3.069×3 | 7.744×2 | 103.13 |
| | | | | | P2$_1$/c | 3.5 | 4.8 | 8.5 | 5.303 | 3.066×3 | 7.725×2 | 103.00 |
| LDH100 | 2.0 | 0.5 | 1.5 | 48.6 | C2/m | 4.5 | 6.4 | 18.3 | 5.313 | 3.071×3 | 7.742 | 103.00 |
| LDH85 | 1.9 | 0.7 | 1.0 | 70.1 | C2/m | 4.72 | 6.47 | 19.15 | 5.279 | 3.058×3 | 7.690 | 103.27 |

**XRD structure analysis**: Diffraction patterns for clays and LDHs are typically dominated by the distance and relative ordering of their layers with high electron density (Figure 5). Carbonate LDH's are typically described in the R-3m spacegroup, which also in the present case allowed to assign the most intense reflections. The weaker reflections and intensity distribution could however not be accounted for. The symmetry operations in R-3m artificially enforce identical, averaged coordination environments of both cations, in this case Zn$^{2+}$ and Al$^{3+}$, which in reality display significantly different bond lengths towards oxygen in octahedral coordination. Further refinement using a trigonal cell (P-3 1 m, $a_t$≈5.3 Å, $c_t$≈22.6 Å) with 3 layers improved the description. P-3 1 m is a subgroup of the R-3m and allows for cation ordering, with $a_t$ representing the distance between second nearest octahedra instead of nearest octahedra. Its observation clearly marks systematic ordering of the cations in the layers. However, the weaker reflections and intensity distribution could still not be accounted for even in this trigonal description. Therefore, the space group was further transformed to its subgroup C2/m using following vector transformation: $\vec{a} = \vec{a_t}, \vec{b} = -\vec{a_t} - 2\vec{b_t}, \vec{c} = -\vec{a_t} - \frac{1}{3}\vec{c_t}$, in full accordance with the single-layer model suggested by Marappa et al.[9] This led to satisfactory descriptions of LDH85 and LDH100. For the structure refinement, a structure model was simulated by FOX for LDH100 in space group C2/m to obtain starting parameters for carbonate and interlayer water. Interestingly, while water molecules were found on very similar positions as reported in literature, the anion position and orientation differed. Carbonate was found slightly tilted in such a way that two of its oxygen atoms were in hydrogen bonding distance to hydroxyls in upper and lower cation layers. Using this model for Rietveld refinement by GSAS led to reasonable description of the powder patterns. Carbonate was treated as rigid body, introduced at the site deduced by FOX and left free to rotation and translation. The tilt of the molecule assumed in the Rietveld analysis was extremely close to the one observed by Monte Carlo simulation with FOX. The water molecules remained close to positions equidistant to upper and lower cation plane and were fixed at this height. Owing to the space group symmetry, 8 carbonate molecules per unit cell are generated on symmetrically equivalent sites. The occupation factor of carbonate refined to a consistent value around ⅛ and was also fixed. Inspection of the resulting model revealed the carbonates roughly are located in rows of 4 molecules each along a, close to y = 0 and 0.5. Within each of these rows only one of the 4 positions can be occupied without steric conflict. Surprisingly, description of LDH125 in the same cell and symmetry failed. This pattern could be indexed only after doubling of the c-axis (C2/m→C2/c), resulting in a 2-layer model and subsequent removal of C-centering. Accounting for the observation by NMR of a single site for Al, space groups P2/n and P2$_1$/c, being primitive subgroups of C2/c, were further investigated. In both space groups, satisfactory refinements with reasonable water and anion distributions were achieved. In both arrangements, the carbonate ions assumed very similar locations and orientations as observed for LDH85 and LDH100 described in the C2/m model, which averages molecule positions over neighboring layers and within them. In the P2$_1$/c structure model, each anion site can be occupied without steric hindrance by other carbonates. Here, all rows found in the C2/m setting of LDH85 and LDH100 are potentially occupied, but only with 1 instead of 4 possible tilts. In neighboring layers, the carbonates are at the same position but rotated around b by 180°. Occupation refined to a value close to 0.5 and then was fixed in agreement with chemical composition and to assure charge neutrality. The resulting carbonate vacancies in this structure model could occur randomly as no steric hindrance occurs. Ordering of the vacancies would result in a structure model almost indistinguishable from the solution in P2/n, explaining probably why both space groups allowed reasonable description of the structure. The water molecules in space group P2$_1$/c were found quite close to the oxygen atoms of CO$_3^{2-}$. The positions were not stable during refinement but moved slightly above and below the center-plane between upper and lower cation-layer, suggesting a possible split position. Hence, the water in each row along a are removed so that molecules were fixed on height 0.75, which led to stable refinement albeit with relatively large thermal factors for water. Occupation of these sites indicates that each vacancy of a carbonate species is occupied by 3 water molecules, because simultaneous presence of water and anion in corresponding sites is sterically impossible.

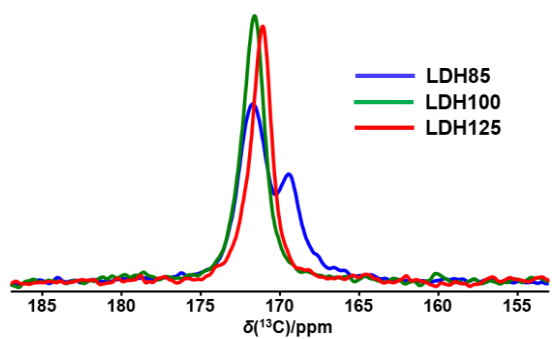

**Figure 4**: $^1$H decoupled $^{13}$C CPMAS spectra (contact time = 1.5 ms) of ZnAl-CO$_3$ LDHs synthesized at different pH recorded at 11.4 T.

Table 2. Al speciation and quantification in ZnAl-CO3 LDHs synthesized under different pH conditions.

| | LDH85 | | LDH100 | | LDH125 | |
|---|---|---|---|---|---|---|
| T site | Fraction (%) | $δ_{iso}$; $C_Q$ (ppm; MHz) | Fraction (%) | $δ_{iso}$; $C_Q$ (ppm; MHz) | Fraction (%) | $δ_{iso}$; $C_Q$ (ppm; MHz) |
| Al-1 | 18.5 | 14.6; 0.73 | 59.8 | 14.6; 0.82 | 65.7 | 14.3; 0.68 |
| Al-2 | 81.5 | 13.3; 3.00 | 40.2 | 11.1; 1.75 | 34.3 | 11.5; 1.85 |

In space group P2/n, a different arrangement is found for carbonate. Here, compared to the C2/m supergroup, alternating rows of carbonate ions are removed in their entirety, so that in each remaining row, two anion positions remain, which cannot be occupied simultaneously. Between layers, these rows strictly alternate so that between Al-sites in neighboring layers there is exactly one carbonate molecule interacting with both. Furthermore, the found orientation suggests the molecule orientation can flip, though except for the thermal factors no further evidence can be given for this assumption. The found occupation of 0.5 can be achieved by removing every second carbonate ion along the rows in $a$-direction, which also removes the steric hindrance. Interestingly, this arrangement of carbonate also results if the vacancies, resulting from an occupation of 50% in spacegroup $P2_1/c$ are ordered. The regions where anion rows are missing are occupied by water molecules, which among themselves and with the carbonate are also in hydrogen bonding distance. Owing to the symmetry of the space group, here the water molecules clearly assume split positions, indicating interaction with hydroxyl groups either above or below according to statistic or dynamic disorder. This hints at order within the rows of carbonate ions along $a$, which is further supported by water positions possibly bridging the ions by hydrogen bonding. Testing if a further reduction of symmetry, i.e. not only strict order within, but also between rows occurs did not lead to any improvement. Which of the two arrangements is realized to which degree in the sample LDH125, at present cannot be clearly decided. Still, it can be assumed based on the similar location and orientation of the anions and water molecules in all three samples, that already in LDH85 and LDH100, regions with similar order exist, even though these are small. Hence, these materials can be satisfactorily described in C2/m, where all sites with the same crystallographic environment with respect to layer atoms are occupied statistically. In the very highly crystalline LDH125, the ordered domains become large enough, to lead to the observed break in symmetry. The refined structures of ZnAl-$CO_3$ LDHs are presented in Figure 5. The structure analysis, revealed similar environment for Al and for carbonate in the crystalline fractions of all 3 samples. The highly symmetric Al-1 position detected by NMR correlates to the fraction of Al present in highly ordered domains described by $P2_1/c$ and P2/n. Based on this, it can be concluded the local ordering in these symmetries is present in all three samples, even though the long-range order, i.e. the size of ordered domains greatly varies, being the largest in sample LDH125 and the smallest in LHD85. Furthermore, not only does the long-range order of the crystalline domains vary, also the amount of material not contributing to Bragg-scattering at all, i.e. X-ray amorphous material increases with lower synthesis pH.

**Advanced NMR spectroscopy**. To further test the derived structure models, LDH85 and LDH125 were prepared with 20% $^{13}C$ enrichment and advanced NMR analysis was performed. The $^1H$-$^{13}C$ CPMAS spectrum for LDH125 (Figure 4, red trace) showed a single, slightly distorted resonance envelope with a maximum at 171 ppm. $^1H$-$^{13}C$ heteronuclear correlation experiments (HETCOR) on the $^{13}C$ enriched LDH125 readily reveal two different carbon species at 170.9 and 171.5 (Figure 6a, top axis), which each correlate to a different proton pool exhibiting a chemical shift between 4 and 5 ppm, respectively (the correlations are shown by green and red lines respectively). Combining the $^1H$-$^{13}C$ HETCOR with $^{13}C$-$^{27}Al$ HETCOR and $^{27}Al$ double quantum - single quantum (DQ-SQ) spectra (Figure 6a, 6b & 6d) provides a clear view on the speciation of carbonate and Al in the sample, and how they correlate. The carbonate species at 170.9 ppm interacts exclusively with Al-1 (green correlation in Figures 6a & 6b; $^{27}Al$ $\delta_{iso}$=14.3 ppm and $C_Q$ = 0.68 MHz, supra), the symmetric local environment of Al surrounded by 6 Zn atoms, described in spacegroups $P2_1/c$ and P2/n. From the $^{27}Al$ DQ-SQ spectrum shown in Figure 6d, it can be seen this LDH derived Al site (Al-1), is fully separated from all other types of Al present in the sample (exclusively self-correlations). These other Al sites are part of a carbonate bearing side-phase and interact with the carbonate species with chemical shift at 171.5 ppm (red and brown correlations in (Figures 6a, 6b & 6d). Comparing these results with corresponding datasets recorded for 20% $^{13}C$ enriched LDH85, the differences between both samples are evident. As expected from the 1D $^{27}Al$ MAS spectra, LDH85 contains a fraction of pure LDH phase (Al-1, green correlations in Figures 6e, 6f & 6h). In addition to this fraction of pure LDH phase, aluminium- and carbonate-containing side-phases represent the majority of this sample. The $^1H$-$^{13}C$ HETCOR spectrum (Figure 6e) indeed reveals two $^{13}C$ resonances not corresponding to pure ZnAl LDH exchanged carbonate. The resonance at 171.9 ppm correlates to $^1H$ species at 5 ppm and to an Al species with a chemical shift at 12.2 ppm (Figure 6h, red correlation). The carbonate with chemical shift at 169.5 ppm correlates not to one, but to three different $^1H$ signals with chemical shifts *viz.* 1.8 (orange line), 5 ppm (red line) and 8 ppm (grey line). These $^{13}C$ - $^1H$ correlations demonstrate those side-phase derived carbonates correlate to different $^1H$ reservoirs as compared to the carbonate contained in the pure LDH phase, proving the side phase is locally apart from the ordered LDH. This also corroborates the interpretation of the CPMAS build up-decay results (supra). In $^{13}C$-$^{27}Al$ HETCOR (figure 6f), the side-phase derived $^{13}C$ resonance at 169.5 ppm correlates to multiple $^{27}Al$ resonances with chemical shifts ranging from 10.5 up to 15 ppm, (Figure 6f, blue, grey, and orange lines correlating to populations at 10.5, 13.8 and 15 ppm), all contained in the envelope of the Al_2 resonance observed in the direct excitation $^{27}Al$ spectrum (Fig 6g inset; $^{27}Al$ $\delta_{iso}$=13.3 ppm and $C_Q$ = 3 MHz). The $^{27}Al$ DQ-SQ spectrum (Fig 6h) shows exclusively autocorrelations, again confirming the LDH and the side-phases exist in fully separate parts of the sample, most likely, the crystalline and amorphous phases observed with SEM and deduced from XRD. From the chemical shifts observed in the $^{13}C$ NMR spectra, also the presence of pure zinc containing sidephases, such as hydrozincite [$Zn_5(CO_3)_2(OH)_6$]

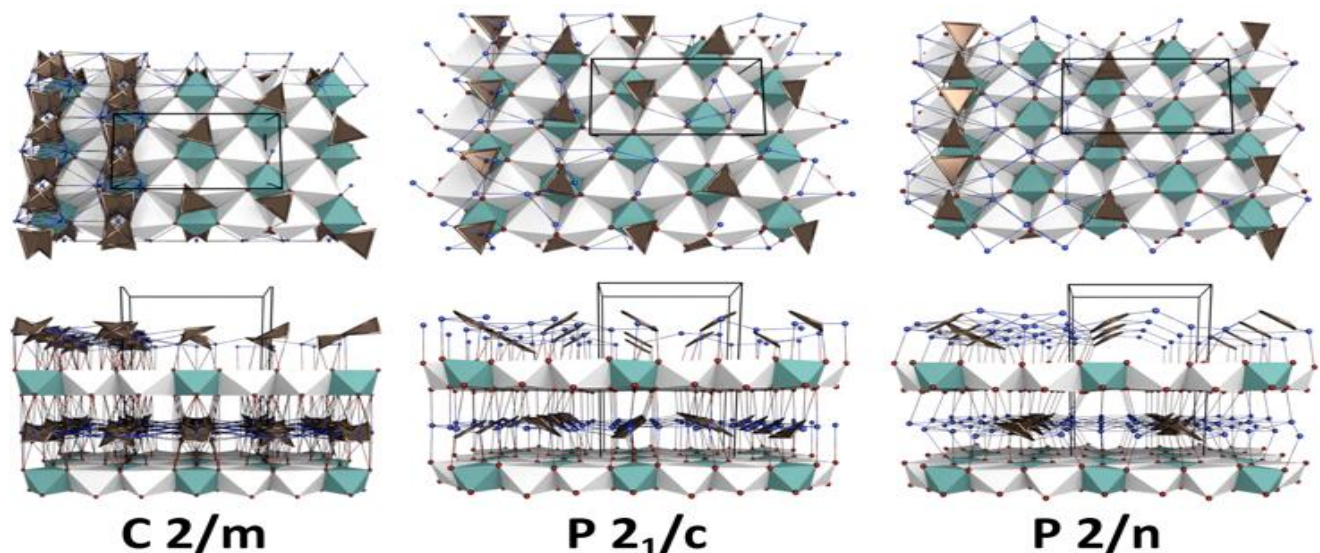

**Figure 5**: Refined structure of LDH85 and LDH100 with space group C2/m and the two arrangements of LDH125 with space groups P2/n and $P2_1/c$. Each structure shows crystallographic sites of carbonate (left) and possible arrangements at occupations of 50%.

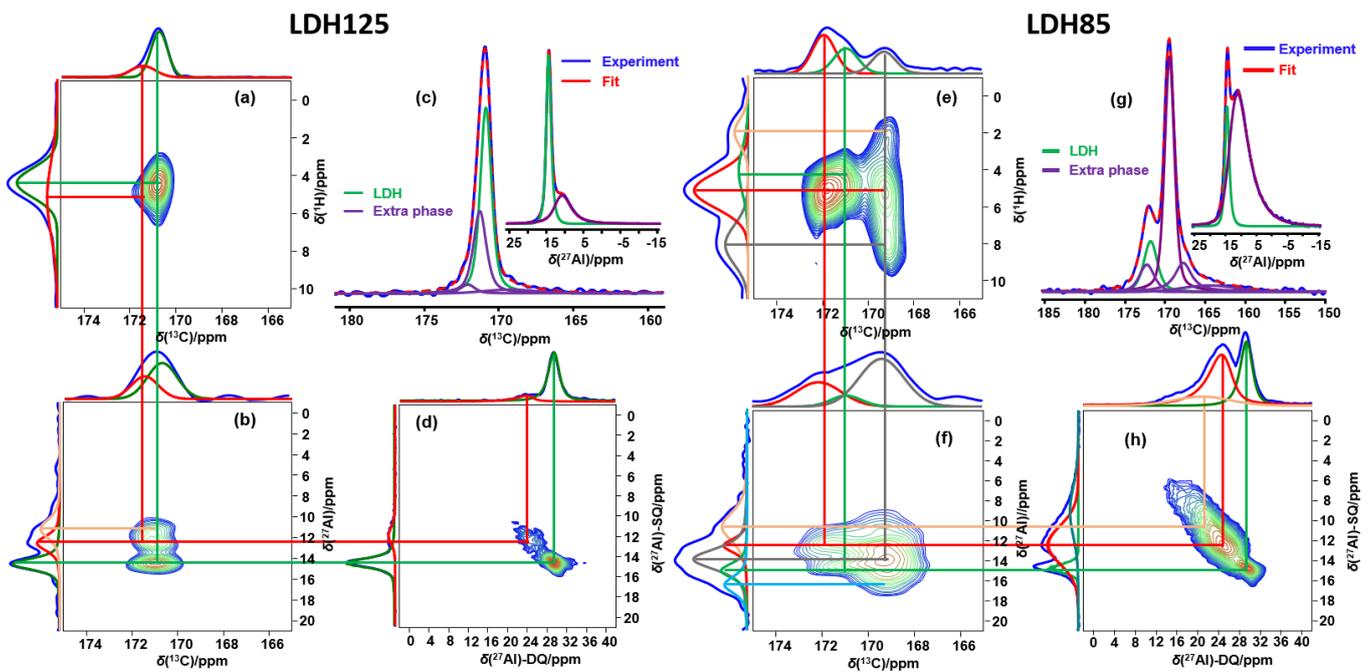

**Figure 6**: $^1H$-$^{13}C$ heteronuclear correlation spectra of LDH125 (a) and LDH85 (e) recorded at 1 ms CP contact time, $^{13}C$-$^{27}Al$ correlation spectra of LDH125 (b) and LDH85 (f) using D-HMQC-IR pulse sequence, $^{27}Al$ DQ-SQ spectra of LDH125 (d) and LDH85(h) and direct excitation $^{13}C$ and $^{27}Al$ (inset) spectra of LDH125 (c) and LDH85 (g). The spectra were measured on 20% $^{13}C$ enriched samples at 18.8T.

can be excluded.[27,28]

The information obtained from the different NMR experiments, the second phase present can be safely attributed to metal oxy-hydroxy carbonate phase. Formation of such an extra phase is entirely possible in the pH ranges employed in the synthesis. To enable quantitative description of the $^{13}C$ speciation in the LDH material, $^1H$ decoupled $^{13}C$ direct excitation spectra of LDH125 and LDH85 with *ca.* 20 % $^{13}C$ enrichment were acquired (Figure 6c and 6g respectively**)**. Considerable difference in contribution of individual resonances can be observed in the direct excitation spectra (Figure 6c and 6g ) compared to the CPMAS spectra (Figure 4), since CPMAS is non-quantitative. Spectral decomposition of the direct excitation spectrum, implementing the information from the different HETCOR experiments, yielded the relative quantifications summarized in table S1. LDH125 contains 3 carbon species with chemical shifts at 170.9 (60 %), 171.5 (28 %) and 169.5 (12 %) ppm respectively (Figure 6c). The signal at 170.9 ppm, dominantly present in LDH125, correlates to the Al resonance at 14.3 ppm associated with the crystalline LDH phase and can indisputably be assigned to $CO_3^{2-}$ ion exchanged onto the LDH. The $^{13}C$ signals at 172.3 and 169.5 ppm correspond either to carbonate species adsorbed onto the extra phase or to surface complexation. It is important to notice the fractions of LDH phase obtained from the decomposition of $^{27}Al$ (62 % LDH fraction) and $^{13}C$ direct excitation spectra (60 % LDH fraction) correspond to a 1:1 ratio. Spectral decomposition of the LDH85 $^{13}C$ direct excitation spectra yielded 5 components assigned to $CO_3^{2-}$ ion exchanged onto the LDH (171.9 ppm, 16 %), and $CO_3^{2-}$ species (172.3 ppm, 10%; 169.5 ppm, 50 %; 168.2 ppm, 16 %) and $HCO_3^-$ species (164.5 ppm, 8 %) adsorbed onto the extra phase (Figure 6g**)**. Also in the case of LDH85, the fraction of ordered LDH in the material obtained from $^{27}Al$ (LDH fraction 16.8 %) and $^{13}C$ (LDH fraction 15.9 %) spectral decomposition correspond well with each other, confirming the assignments.

As neither LDH125 nor LDH100 contain $HCO_3^-$ species in the crystalline fraction of the samples, LDHs synthesized in carbonate buffers at pH 8.5, 10 and 12.5 respectively, must exhibit an extremely high selectivity for divalent carbonate anions. This extreme selectivity for $CO_3^{2-}$ was confirmed by equilibrating the LDH125 aqueous carbonate solutions at pH 11.2, 9.8 and 8.4, varying the $CO_3^{2-}/HCO_3^-$ ratio between 1/0, 0.5/0.5 and 0/1. When LDH125 was equilibrated in 100 % $Na_2CO_3$, 100 % $NaHCO_3$ or 50/50 $Na_2CO_3$-$NaHCO_3$ buffer solutions and subsequently dried, no differences were observed in either the $^{27}Al$ or $^{13}C$ NMR spectra (Figure S-3), confirming the exclusive adsorption of carbonate and thus also extreme selectivity towards the divalent carbonate anion. Such selectivity can only be derived from ordering of Al centers providing a local charge density that can optimally be compensated by divalent anions.

## Conclusion

ZnAl-$CO_3$ LDHs were synthesized at different pH (pH 8.5, pH 10.0, pH 12.5). It was clear that increasing synthesis pH positively impacted LDH structure and morphology with a long range ordered crystalline LDH structure synthesized at pH 12.5. $^{27}Al$ MAS NMR indicated that all LDH samples contained two different six-coordinated Al local environments. Enhanced LDH crystallinity at higher pH also induced higher relative quantification of local Al order surrounded by 6 Zn atoms. The quantification of carbonate species in LDHs synthesized at high pH (pH 12.5) showed the presence of one type of carbonate interlayer anion in the crystalline fraction of all samples. Al-C correlation experiments revealed high selectivity for divalent carbonate anions in the LDH phase. For lower synthesis pH (LDH85, pH 8.5), different CP build-up behaviors proved the existence of different carbonate species, associated to crystalline and disordered phase fractions. The amorphous phase was found to be less selective and contains isolated bicarbonate ions next to carbonate anions. $^{13}C$ relative amounts based on spectral decomposition of $^{13}C$ single pulse spectra corresponded well with the $^{27}Al$ relative distribution, confirming the assignment to LDH and the amorphous environments. Monoclinic single-layer symmetry models with space group C2/m could describe crystal structure of LDHs synthesized at lower pH, but resulted in a superposition of several possible carbonate orientations. For the hypercrystalline LDH structure synthesized at pH 12.5, however, a two-layer model with either P2/n or P2$_1$/c space group, revealed anion ordering. The observed anion order also could be deduced to be present by merging insights from NMR and XRD, indicating that the failure to detect it by diffraction in LDH85 and LDH100 entirely is caused by the decreasing size of ordered domains in the crystalline fraction of the materials.

## Experimental Section

**LDH synthesis and basic characterization**

ZnAl-$CO_3$ LDHs were synthesized at three different pHs. Samples prepared at pH 8.5, pH 10.0 and pH 12.5 will be referenced to as LDH85, LDH100 and LDH125, respectively. LDHs were synthesized by dosing a pre-mixed $Zn(NO_3)_2$-$Al(NO_3)_3$ solution (50 mL; Zn/Al ratio = 2; total metal concentration = 1 mol/L) at a constant rate of 12.5 mL/h into a $Na_2CO_3$ buffer solution (200 mL; 0.1 mol/L) stated at pH 8.5, pH 10.0 or pH 12.5, respectively, by continuous titration with NaOH (1 mol/L). Synthesis suspensions were aged statically at 60 °C for 24 h, followed by centrifugation and a rinsing step with demineralized water. The resulting solid phase was subsequently dried at 60 °C. Dried LDH125 was exchanged for 24 h in 100% $Na_2CO_3$, 100% $NaHCO_3$ and 50/50 $Na_2CO_3$-$NaHCO_3$ aqueous solutions (0.5 g LDH in 50 mL; 0.1 mol/L). $^{13}C$ enriched-LDH85 and -LDH125 were synthesized in a buffer with 20% $^{13}C$ abundance. Scanning electron microscopy (SEM) using a FEI-Nova Nano-SEM 450 resolved LDH crystal morphology. The Zn and Al chemical composition of the LDH samples were determined on an axial simultaneous ICP-OES (Varian 720-ES) with cooled cone interface and oxygen free optics where the samples were introduced into the machine via a Varian SPS3 Sample Preparation System. Carbon analysis of the different LDH phases was performed by dry flash combustion on a EA1108 apparatus (Carlo Erba). Thermal gravimetric analysis (TGA) was performed on a TGA Q500 (TA instruments, Belgium) under nitrogen flow at a heating rate of 5 °C/min until 900 °C. Nitrogen adsorption isotherms were measured at -196 °C using a Quantachrome Autosorb-1 instrument to calculate surface area and porosity of the LDH materials. The samples were degassed at 150 °C prior to the measurement. The BET surface area was estimated in the P/P0 range of 0.05 – 0.3, using the best linear fit.

Advanced NMR spectroscopy

$^1H$, $^{27}Al$ and $^{13}C$ NMR spectra were recorded at Larmor Frequencies of 500.87, 130.52 and 125.96 MHz respectively on a Bruker 500 MHz spectrometer equipped with 4 mm H/X/Y MAS probe. Samples were filled in a 4 mm $ZrO_2$ rotor and spun at 15 kHz for direct excitation $^1H$ and $^{27}Al$ MAS NMR and at 10 kHz for $^{13}C$ MAS NMR measurements, respectively. $^1H$ spectra were recorded using a π/2 pulse at 83 kHz, with the probe matched to a constant Q-factor of 600 to allow absolute quantification,[29] 3 s recycle delay and 64 transients. $^1H$ NMR spectra were referenced to TMS (0.00 ppm), using adamantane at 1.81 ppm as a secondary reference. $^1H$ decoupled $^{27}Al$ spectra were recorded using a π/12 pulse at 87 kHz, 2 s recycle delay, $^1H$ SPINAL-64 decoupling[30] and accumulation of 1024 transients. $^{27}Al$ spectra were referenced to 0.5 M aluminum nitrate solution at 0.00 ppm. Spectral decomposition was performed using DMFIT software.[31] $^1H$-$^{13}C$ cross-polarization magic angle spinning (CPMAS) spectra were acquired at 10 kHz MAS, with 72.5 kHz RF field on $^{13}C$, a shaped pulse on $^1H$ (100-70 %) for the contact, 1024 scans, recycle delay of 3 s and $^1H$ SPINAL-64 decoupling.[30] Different contact times were employed to investigate cross-polarization buildup and decay behavior. For the $^{13}C$ enriched-LDH samples, $^1H$ decoupled $^{13}C$ spectra were acquired using a π/2 pulse at 73.5 kHz, 1200 s recycle delay, $^1H$ SPINAL-64 decoupling at 60 kHz and accumulation of 64 transients.

$^1H$-$^{13}C$, $^{27}Al$-$^{13}C$ hetero nuclear correlations and $^{27}Al$-$^{27}Al$ homonuclear correlation experiments were recorded on a 18.8 T standard bore spectrometer equipped with a Bruker Avance NEO console. $^{27}Al$-$^{13}C$ hetero-nuclear correlation were recorded on a 3.2 mm HX Bruker probe with a frequency splitter (REDORBOX from NMR service) that allows to simultaneously tune the X channel to the Larmor frequencies of $^{13}C$ and $^{27}Al$.[32–35] The sample was spun at 20 kHz. Correlations were obtained by a D-HMQC sequence[36–38] using SFAM-1 dipolar recoupling on the indirect dimension channel with SFAM maximum offset of 60 kHz and SFAM maximum RF-field of 65 kHz. Central Transition (CT) selective of quadrupolar $^{27}Al$ spin 90° and 180° pulses of 14 µs and 28 µs respectively were used (corresponding to a liquid RF-field of 6 kHz) and $^{13}C$ 90° pulse were 4 µs long. Spinal64 $^1H$ decoupling was used during SFAM recoupling periods using 4.2 µs pulses at RF-field of 120 kHz. A hyperbolic secant pulse of 4 ms at 11 kHz maximum RF-field using one offset swept from 210 to 190 kHz allowed to enhance the $^{27}Al$ CT signal by a factor of about 2. For LDH85, 2D acquisition was using the States-TPPI scheme with 134 rows recorded and a rotor synchronized indirect spectral window of 20 kHz. The recycling delay was 1.5 s and 256 scans were averaged during 5 ms. Experimental time was 10 h. For LDH125, 2D acquisition was using the States-TPPI scheme with 98 rows recorded and an indirect spectral window of 10 kHz. Recycling delay was 3 s and 736 scans were averaged, SFAM recoupling was applied during 2.5 ms. The total experimental time was 60 h. $^{27}Al$-$^{27}Al$ homonuclear correlations were recorded using a 3.2 mm HX probe at 20 kHz MAS. Homonuclear recoupling was performed using the $BR2_2^1$ scheme[39] with an RF-field of about 3.3 kHz (effective CT selective RF-field of 10 kHz or half the spinning speed) and a recoupling period of 600 µs was used. Offset was set slightly off resonance at about 24 ppm. For 2D spectra, 80 rows were recorded using the States-TPPI scheme with $t_1$ increment of 50 µs, 64 (LDH125) or 128 (LDH85) scans and recycling delay 1 s. Experimental time was 1 h 30 min. Signal was enhanced by a factor of about 2 by a Quadruple Frequency Sweep[40] of 1 ms long with frequency sweeping by 20 kHz at offsets ±100 and ±200 kHz and 15 kHz RF-field. $^1H$-$^{13}C$-HETCOR spectra were recorded on a 1.3 mm HX probe at 60 kHz MAS. CPMAS was using 1 ms contact time with RF-field on $^1H$ and $^{13}C$ of c.a. 210 kHz and 150 kHz respectively using a ramp between 90 and 100 % on $^1H$. 16 scans were averaged using 2 s recycling delay. The 2D spectra were constructed using the States-TPPI hypercomplex scheme with F1 spectral window of 60 kHz, recording 56 (LDH85) or 96 (LDH125) rows leading to an experimental time of 30 to 50 minutes. A $SW_f$-TPPM[41] $^1H$ decoupling was used during acquisition with an RF-field of 15 kHz and a base pulse length of 33 µs being swept from 0.78 to 1.22 with phase alternation of ±12.5°.

XRD structure analysis

XRD patterns were recorded on a STOE STADI MP diffractometer with focusing Ge(111) monochromator (Cu $K_{α1}$ radiation) in Debye−Scherrer geometry with a linear position sensitive detector (PSD) (6° 2θ window) with a step width of 0.5 °2Θ and internal resolution of 0.01 °2Θ. Diffractograms initially were indexed using the Visser and Louër algorithms.[42,43] Potential space groups were determined based on the cation ordering reported by Marappa et. al.[9] Candidate space groups, all being C2/m or subgroups thereof, were evaluated using the FOX software,[44] which also served to generate structure models for Rietveld refinement with the GSAS program suite. Refinement occurred jointly with the NMR investigation, to assure the structural description is in agreement with all experimental results .[45,46]

## Acknowledgements

This research was funded by a Ph.D. grant (K.L.) from the Agency for Innovation by Science and Technology (IWT). This work was supported by the Hercules Foundation (AKUL/13/21). NMRCoRe acknowledges the Flemish government, department EWI for infrastructure investment via the Hermes Fund (AH.2016.134) and for financial support as International Research Infrastructure (I001321N: Nuclear Magnetic Resonance Spectroscopy Platform for Molecular Water Research). J. A. M. acknowledge the Flemish Government for long-term structural funding (Methusalem). This project has received funding from the European Union's Horizon 2020 research and innovation programme under grant agreements No 731019 (EUSMI) and No 834134 (WATUSO).

**Keywords:** NMR Crystallography • Layered double hydroxide (LDH) • Solid-state NMR • Rietveld refinement • keyword 5

## References

[1]  X. Duan, D. G. Evans, *Layered Double Hydroxides*, Springer Berlin Heidelberg, **2006**.

[2] J. J. Bravo-Suárez, E. A. Páez-Mozo, S. T. Oyama, *Quim. Nov.* **2004**, *27*, 601–614.

[3] A. F. Morais, I. G. N. Silva, S. P. Sree, F. M. de Melo, G. Brabants, H. F. Brito, J. A. Martens, H. E. Toma, C. E. A. Kirschhock, E. Breynaert, D. Mustafa, *Chem. Commun.* **2017**, *53*, 7341–7344.

[4] A. F. Morais, F. O. Machado, A. C. Teixeira, I. G. N. Silva, E. Breynaert, D. Mustafa, *J. Alloys Compd.* **2019**, *771*, 578–583.

[5] I. G. Richardson, *Acta Cryst.* **2013**, *B69*, 150–162.

[6] S. V. Krivovichev, V. N. Yakovenchuk, E. S. Zhitova, A. A. Zolotarev, Y. A. Pakhomovsky, G. Y. Ivanyuk, *Mineral. Mag.* **2010**, *74*, 833–840.

[7] K. Jayanthi, S. Nagendran, P. V. Kamath, *Inorg. Chem.* **2015**, *54*, 8388–8395.

[8] L. Pachayappan, S. Nagendran, P. V. Kamath, *Cryst. Growth Des.* **2017**, *17*, 2536–2543.

[9] S. Marappa, P. V. Kamath, *Ind. Eng. Chem. Res.* **2015**, *54*, 11075−11079.

[10] S. S. C. Pushparaj, C. Forano, V. Prevot, A. S. Lipton, G. J. Rees, J. V Hanna, U. G. Nielsen, *J. Phys. Chem. C* **2015**, *119*, 27695–27707.

[11] S. S. C. Pushparaj, N. D. Jensen, C. Forano, G. J. Rees, V. Prevot, J. V Hanna, D. B. Ravnsbæk, M. Bjerring, U. G. Nielsen, S. Shiv, C. Pushparaj, N. D. Jensen, C. Forano, G. J. Rees, V. Prevot, J. V Hanna, D. B. Ravnsbæk, M. Bjerring, U. G. Nielsen, *Inorg. Chem.* **2016**, *55*, 9306–9315.

[12] P. J. Sideris, U. G. Nielsen, Z. Gan, C. P. Grey, *Science* **2008**, *321*, 113–117.

[13] S. Cadars, G. Layrac, C. Gérardin, M. Deschamps, J. R. Yates, D. Tichit, D. Massiot, *Chem. Mater.* **2011**, *23*, 2821–2831.

[14] P. J. Sideris, F. Blanc, Z. Gan, C. P. Grey, *Chem. Mater.* **2012**, *24*, 2449–2461.

[15] P. Sahoo, S. Ishihara, K. Yamada, K. Deguchi, S. Ohki, M. Tansho, T. Shimizu, N. Eisaku, R. Sasai, J. Labuta, D. Ishikawa, J. P. Hill, K. Ariga, B. P. Bastakoti, Y. Yamauchi, N. Iyi, *ACS Appl. Mater. Interfaces* **2014**, *6*, 18352–18359.

[16] Y. Kuroda, Y. Miyamoto, M. Hibino, K. Yamaguchi, N. Mizuno, *Chem. Mater.* **2013**, *25*, 2291–2296.

[17] A. Di Bitetto, G. Kervern, E. André, P. Durand, C. Carteret, *J. Phys. Chem. C* **2017**, *121*, 6104–6112.

[18] F. L. Theiss, G. A. Ayoko, R. L. Frost, *Appl. Surf. Sci.* **2016**, *383*, 200–213.

[19] X. Sun, S. K. Dey, *J. Colloid Interface Sci.* **2015**, *458*, 160–168.

[20] J. T. Kloprogge, L. Hickey, R. L. Frost, *J. Solid State Chem.* **2004**, *177*, 4047–4057.

[21] S. Cousy, N. Gorodylova, L. Svoboda, J. Zelenka, *Chem. Pap.* **2017**, *71*, 2325–2334.

[22] D. T. Y. Chen, *Can. J. Chem.* **1973**, *51*, 3528–3533.

[23] S. P. Newman, W. Jones, in *Supramol. Organ. Mater. Des.*, Cambridge University Press, **2009**, pp. 295–331.

[24] J. B. d'Espinose de Lacaillerie, C. Fretigny, D. Massiot, *J. Magn. Reson.* **2008**, *192*, 244–251.

[25] G. Czjzek, J. Fink, F. Götz, H. Schmidt, J. M. D. Coey, J. P. Rebouillat, A. Liénard, *Phys. Rev. B* **1981**, *23*, 2513–2530.

[26] C. Vinod Chandran, C. E. A. Kirschhock, S. Radhakrishnan, F. Taulelle, J. A. Martens, E. Breynaert, *Chem. Soc. Rev.* **2019**, *48*, 134–156.

[27] G. De Giudici, F. Podda, R. Sanna, E. Musu, R. Tombolini, C. Cannas, A. Musinu, M. Casu, *Am. Mineral.* **2009**, *94*, 1698–1706.

[28] R. Sanna, G. De Giudici, A. M. Scorciapino, C. Floris, M. Casu, *Am. Mineral.* **2013**, *98*, 1219–1226.

[29] M. Houlleberghs, A. Hoffmann, D. Dom, C. E. A. Kirschhock, F. Taulelle, J. A. Martens, E. Breynaert, *Anal. Chem.* **2017**, *89*, 6940–6943.

[30] B. M. Fung, A. K. Khitrin, K. Ermolaev, *J. Magn. Reson.* **2000**, *142*, 97–101.

[31] D. Massiot, F. Fayon, M. Capron, I. King, S. Le Calvé, B. Alonso, J.-O. Durand, B. Bujoli, Z. Gan, G. Hoatson, *Magn. Reson. Chem.* **2002**, *40*, 70–76.

[32] F. Dé Rique Pourpoint, A. Sofia, L. Thankamony, C. Volkringer, T. Loiseau, J. Tré, F. Aussenac, D. Carnevale, G. Bodenhausen, H. Vezin, O. Lafon, J.-P. Amoureux, *Chem. Commun* **2014**, *50*, 933.

[33] S. Li, F. Pourpoint, J. Trébosc, L. Zhou, O. Lafon, M. Shen, A. Zheng, Q. Wang, J.-P. Amoureux, F. Deng, *J. Phys. Chem. Lett.* **2014**, *5*, 3068–3072.

[34] F. Pourpoint, Y. Morin, R. M. Gauvin, J. Trébosc, F. Capet, O. Lafon, J.-P. Amoureux, *J. Phys. Chem. C* **2013**, *117*, 18091–18099.

[35] F. Pourpoint, J. Trébosc, R. M. Gauvin, Q. Wang, O. Lafon, F. Deng, J.-P. Amoureux, *ChemPhysChem* **2012**, *13*, 3605–3615.

[36] J. Trebosc, B. Hu, J. P. Amoureux, Z. Gan, *J. Magn. Reson.* **2007**, *186*, 220–227.

[37] Z. Gan, *J. Magn. Reson.* **2007**, *184*, 39–43.

[38] B. Hu, J. Trébosc, J. P. Amoureux, *J. Magn. Reson.* **2008**, *192*, 112–122.

[39] Q. Wang, B. Hu, O. Lafon, J. Trébosc, F. Deng, J. P. Amoureux, *J. Magn. Reson.* **2009**, *200*, 251–260.

[40] Q. Wang, J. Trébosc, Y. Li, O. Lafon, S. Xin, J. Xu, B. Hu, N. Feng, J.-P. Amoureux, F. Deng, *J. Magn. Reson.* **2018**, *293*, 92–103.

[41] C. V. Chandran, P. K. Madhu, D. Kurur, **2008**, 943–947.

[42] D. Louër, A. Boultif, *Zeitschrift fur Krist. Suppl.* **2007**, *26*, 191–196.

[43] J. W. Visser, *J. Appl. Crystallogr.* **1969**, *2*, 89–95.

[44] V. Favre-Nicolin, R. Černý, *J. Appl. Crystallogr.* **2002**, *35*, 734–743.

[45] A. C. Larson, R. B. Von Dreele, *Los Alamos Natl. Lab. Rep.* **1994**, *LAUR 86-74*.

[46] B. H. Toby, *J. Appl. Crystallogr.* **2001**, *34*, 210–213.

8